**TITLE**

An Automated Real-Time Approach for Image Processing and Segmentation of Fluoroscopic Images and Videos Using a Single Deep Learning Network

**AUTHORS**

Viet Dung Nguyen, Ph.D., University of Tennessee

Michael T. LaCour, Ph.D., University of Tennessee

Richard D. Komistek, Ph.D., University of Tennessee

**CORRESPONDING AUTHOR**

Viet Dung Nguyen

The University of Tennessee

310 Perkins Hall

1506 Middle Drive

Knoxville, TN 37916

vnguye28@vols.utk.edu

**WORD COUNT**

Abstract : 199

Main text (Excluded references) : 4140

**ABSTRACT**

Image segmentation in total knee arthroplasty is crucial for precise preoperative planning and accurate implant positioning, leading to improved surgical outcomes and patient satisfaction. The biggest challenges of image segmentation in total knee arthroplasty include accurately delineating complex anatomical structures, dealing with image artifacts and noise, and developing robust algorithms that can handle anatomical variations and pathologies commonly encountered in patients.

The potential of using machine learning for image segmentation in total knee arthroplasty lies in its ability to improve segmentation accuracy, automate the process, and provide real-time assistance to surgeons, leading to enhanced surgical planning, implant placement, and patient outcomes. This paper proposes a methodology to use deep learning for a robust and real-time total knee arthroplasty image segmentation.

The deep learning model, trained on a large dataset, demonstrates outstanding performance in accurately segmenting both the implanted femur and tibia, achieving an impressive mean Average Precision (mAP) of 88.83 when compared to the ground truth, while also achieving a real-time segmented speed of 20 frames per second (fps).

We have introduced a novel methodology for segmenting implanted knee fluoroscopic or x-ray images, which showcases remarkable levels of accuracy and speed, paving the way for various potential extended applications.

# 1 Introduction

Total knee arthroplasty (TKA) is a highly effective surgical intervention for osteoarthritis [1]. The success of TKA relies heavily on meticulous pre-operative planning and subsequent evaluation of post-operative outcomes, which are often done through weight bearing kinematics analyses using validated 3D-to-2D X-Ray registration techniques [2]. Unfortunately, these analyses can be time-consuming, labor-intensive, and require substantial amounts of image processing to fully capture the implant details. Processing of the X-Ray images is currently typically done manually due to the presence of noise and the need for expert interpretation in medical imaging [3]. Conventional image processing methods such as filtering, segmentation, feature extraction, registration, and image reconstruction have limitations when applied to TKA videos and images, particularly fluoroscopic images with low resolution caused by reduced radiation dosage. Therefore, the development of a robust, efficient, and fast image processing tool to extract crucial information from these X-Ray images becomes an essential step for efficiently improving the outcomes TKA.

Image segmentation is the image processing method of dividing an image into multiple parts or regions based on certain characteristics such as color, texture, or shape [4]. By segmenting an image into different objects, it becomes easier for a computer algorithm to recognize and track those objects in subsequent frames of a video or image sequence. Additionally, for the 3D-to-2D registration process used in postoperative TKA analyses, the raw fluoroscopy image contains unnecessary information such as bone texture, soft tissue texture, and generic excess image noises. Therefore, it is important to remove this noise and isolate only the information of the implanted components, typically the femur and tibia, which can then be segmented and used for manual or automated registration purposes, as demonstrated in Figure 1.

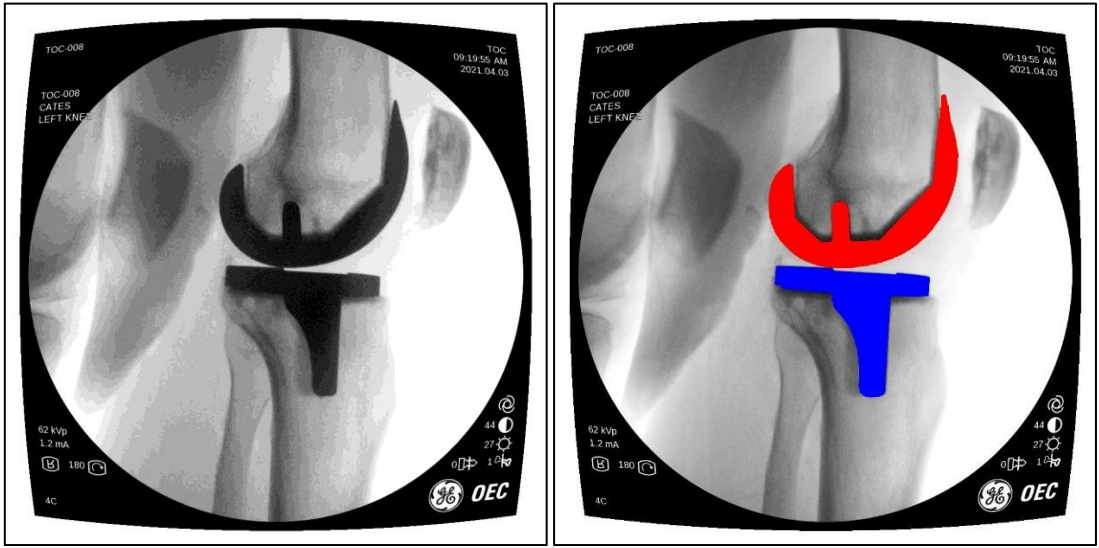

Figure 1. Example of the implanted knee fluoroscopic image (left) and its segmentation with femur (red) and tibia (blue).

The traditional approach to image segmentation involves manual tracing of regions of interest by an expert, which can be time-consuming. While there are several conventional techniques for automated image segmentation, including thresholding [5], edge detection [6], clustering, or statistical shape method [7], these may have a limited accuracy to due to blurry or low-quality fluoroscopic images. Fortunately, with the continual increase of computational power, machine learning-based methods can use artificial intelligence algorithms efficiently to learn how to segment images based on training data. Deep learning is the most widely used machine learning approach for image segmentation, and by incorporating large sets of data accumulated over numerous years of fluoroscopic analyses, deep learning emerges as a highly promising solution for addressing the challenge of automated image segmentation.

Thus, the primary aim of this research is to leverage the power of deep learning in order to develop a real-time, robust, and efficient method for automatically segmenting total knee arthroplasty X-ray or fluoroscopic images.

**2 Method**

The machine learning-based approach proposed for this research consists of three primary subsections: data preparation, training model selection, and model application.

*2.1 Data Preparation: Convert Registration Data to Segmentation Data*

The fluoroscopic database used in this research consists of more than 5,000 fluoroscopic knee images, which have been further associated with 3D implanted models that were manually registered onto their fluoroscopy images. This comprehensive dataset comprises 939 patient records, encompassing a diverse range of implant types such as PS and CR. These records originate from multiple reputable manufacturers, ensuring a rich and varied pool of information. This provides a reliable dataset for a validated 3D-2D registration technique [2] that is aimed at matching the projection of the 3D model with the silhouette of the implant in the 2D image. By using the voxelization projection framework shown in Figure 2, the known registration transformations can produce segmented images of both femur and tibia directly from the registered models. This process generates a raw segmentation dataset consisting of 5,000 pairs of the original fluoroscopic images and corresponding segmented and labeled images of either femur or tibia.

From here, by converting this dataset to the COCO (Common Objects in Context) format [8], which is standard for image processing, it is possible to not only train an appropriate leaning model but also quantifiably compare and verify the accuracy and performance of the proposed deep learning model against other existing methods.

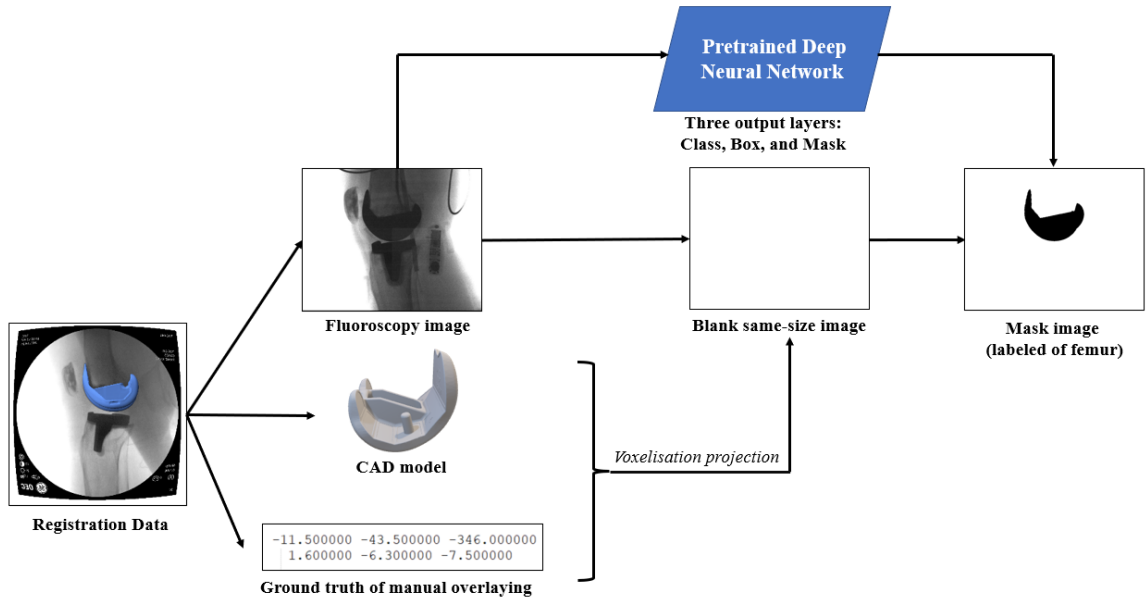

Figure 2. Data preparation of one subject for training the deep neural network of image segmentation.

## 2.2 Training Model Selection and Development

Among advanced deep neural networks available for this task, such as FCN [9], CRF-RNN [10], DPN [11], GCN [12], and Yolact [13], the Yolact algorithm was chosen for its high accuracy and speed, making it a logical choice for knee image segmentation.

Specifically, Yolact is built based on the ResNet model [14], which boasts excellent speeds, around 5 ms to evaluate a mask, so this model proves capable of rapidly segmenting multiple frames in a video. Additionally, the quality and robustness of Yolact has been validated with higher accuracy in comparison with other methods using COCO's dataset [15] with 2.5 million labeled subjects of 91 different objects in 328 thousand images [15].

The TKA segmentation model, based on Yolact, undergoes pruning [16] to match the dimensions of the training dataset. The model takes original fluoroscopic images as inputs and generates three layers of outputs representing labels, boxes, and masks (as depicted in Figure 2). The training data consists of 4,500 subjects, each containing both femur and tibia data. The training process employs the backpropagation method, enhanced by GPU acceleration for faster computation. This procedure is performed on the Alienware Aurora R13 Desktop, equipped with 32 Gb RAM and NVIDIA GeForce RTX 3080 GPU. It runs for 1 million epochs with a batch size of 8 and takes approximately 3 days to complete the training process.

*2.3 Model Applications: Regular Cases, Challenging Cases, and Evaluation*

Upon completion of training the deep neural network, the model was evaluated on a testing set comprising 500 subjects containing both regular images and challenging cases, and subsequently validated in the following sections. The trained model was employed to predict the mask images of the femur and tibia in the implanted knee image. The DNN, after receiving the fluoroscopy image as input, outputs a segmented mask image. This DNN has the ability to predict both femur and tibia simultaneously, and the predicted mask images can be compared to the ground truth in the testing set and also used to assess the accuracy of the trained model in comparison with other segmentation or learning methods.

Due to the dynamic nature of patient data collection, fluoroscopic videos and images used for 3D-to-2D registration can pose various challenges and result in blurry images. For example, Figure 3 (a-i) illustrates several typical worst-case scenarios that can occur during data collection. The first case (a) is characterized by blurry fluoroscopic images due to the low frame rate used by the fluoroscopy machine, done to minimize exposure and to ensure patient and technician safety. Cases (b) and (c) occur when the knee is outside the captured range of the small fluoroscopic machine. Cases (d) and (e) arise when the implanted femur, implanted tibia, and occasionally the resurfaced patella overlap, making it challenging to segment the metal silhouettes into different entities. The next case (f) occurs when both legs have been implanted, yielding additional silhouettes for an algorithm to potentially detect. In case (g), the fluoroscopic data contains noise from a sub-screen inside the image that may affect the model. Finally, the most challenging cases (h) and (i) combine all the above issues, such as blurry edges and overlapping images. Usually, traditional approaches require careful manual oversight in cases such as these, so accurately resolving these cases automatically will be an important foundation for the next sections.

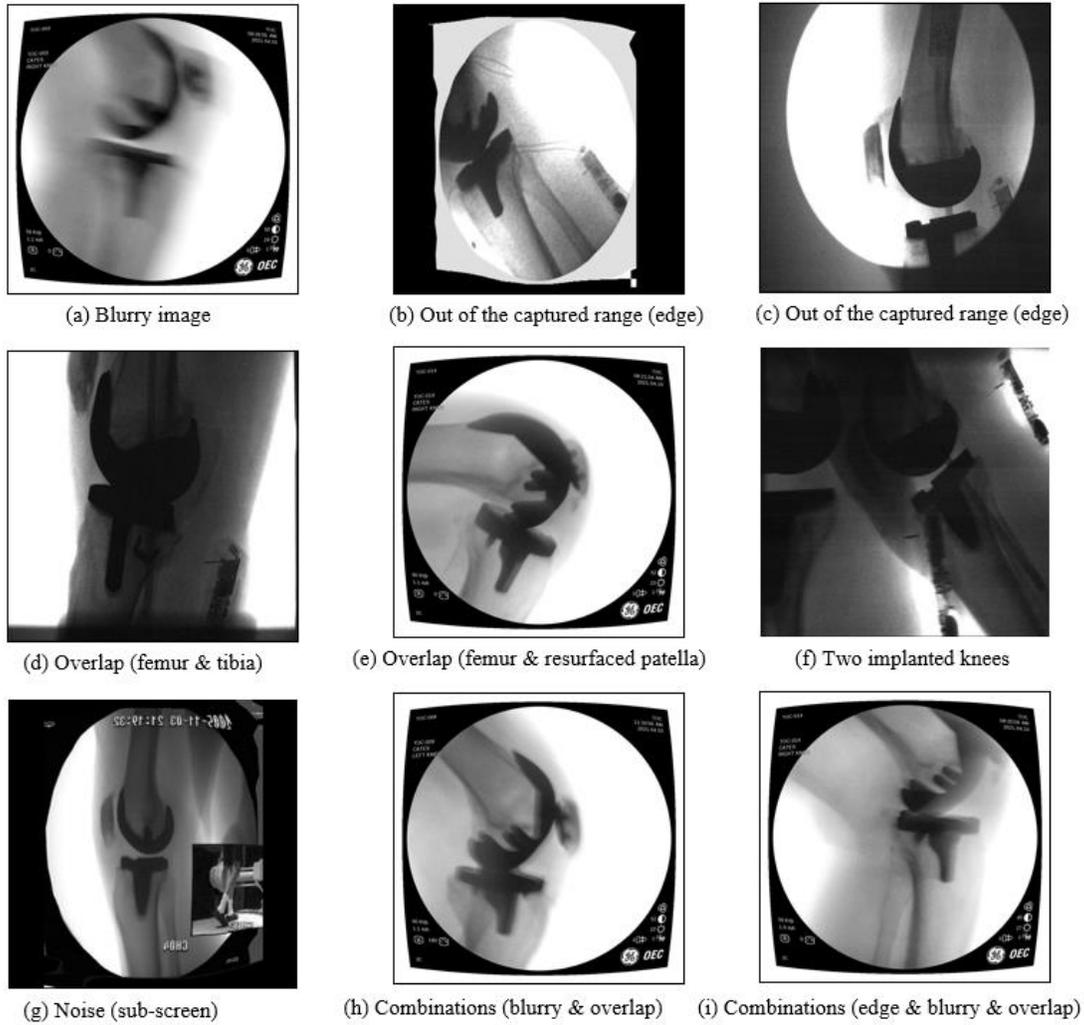

Figure 3. Examples of several challenging cases of implanted knee image segmentation that might be impossible by manual human vision.

## *2.4 Model Evaluation*

The image segmentation algorithm was evaluated based on several metrics to measure overall accuracy and effectiveness. Common metrics used for evaluating image segmentation include Intersection-over-Union (IoU), precision and recall, F1 score, and mean-Average-Precision (mAP). Among those, Mean Average Precision (mAP) is the average of the Average Precision (AP) values across multiple classes or categories [17], which was chosen as the evaluation metric in this research. IoU is used as a threshold to determine whether a predicted bounding box or segmented region is considered a true positive or a false positive. By comparing the IoU between the predicted and ground truth regions against a specified threshold (usually 0.5 or higher), it is determined whether the prediction is considered a correct detection. mAP is then calculated by taking the average precision (AP), as shown in Equation (1), across multiple IoU thresholds. The precision-recall curve is generated by varying the IoU threshold and calculating precision and recall values at each threshold. The AP at each IoU threshold is computed as the area under the precision-recall curve for that threshold. By averaging the AP values across multiple IoU thresholds (e.g., 0.5, 0.75, 0.95, etc.), mAP provides a comprehensive evaluation of the object detection or segmentation model's performance across different levels of overlap between predicted and ground truth regions.

$$mAP = \frac{1}{N}\sum_{k=1}^{N} AP_k \qquad (1)$$

The mAP provides an overall measure of the performance of the system across all categories and is a widely used metric for evaluating object detection models. The mAP metric is popular because it's simple to compare the performance of different models using only a single number. A segmentation model with a higher mAP value indicates greater accuracy. The best three deep learning models for segmenting the commonly referenced

COCO test-dev dataset [15] are Mask R-CNN with mAP=34.7 [18], PA-Net with mAP=36.6 [19], and MS R-CNN with mAP =38.3 [20].

**3 Results**

This section describes the evaluation of the proposed deep learning method in comparison to the other methods as well as the robustness of the deep learning method by segmenting a fluoroscopic video. The 5,000 implanted knee images in the database yield approximately 10,000 known segmented images of both femur and tibia, which were separated into a training set (90%) and testing set (10%). The simulation results of the trained deep learning model revealed impressive accuracy via comparison of the predicted segmentation images with the ground truth using mAP criteria (a metric for evaluating object detection models with a range of [0; 100]) [17].

*3.1 Comparison with the Ground Truth*

Evaluation of the deep learning model was performed on 500 testing images with the known ground truths. Segmenting the mask provides pixel-level delineation of object boundaries, while segmenting the 2D box represents objects using rectangular regions. The mAP for the segmented mask images of the implanted knee by deep learning compared to the ground truth was 88.83, while this number of predicting the 2D box was 97.57, demonstrating high accuracy of the deep learning model for segmenting the femur images compared to the ground truth.

Table 1 presents a comprehensive analysis of the mAP for the predicted registered knee image compared to the ground truth, considering various IoU thresholds. At every IoU threshold, the AP is determined by calculating the area under the precision-recall curve for that specific threshold, thereby establishing the dependence of mAP on IoU. Ranging from 0.50 to 0.95, the IoU thresholds reveal intriguing insights. The "box" metric consistently achieves a perfect 100.0% mAP for IoU thresholds of 0.50 to 0.70, signifying precise alignment between predicted and ground truth bounding boxes. However, for higher IoU thresholds, accuracy slightly declines, ranging from 98.96% to 80.33%. Similarly, the "mask" metric demonstrates an outstanding 100.0% mAP for lower IoU thresholds, gradually decreasing from 99.35% to 86.76% as thresholds increase. Although the predicted registered knee image performs exceptionally well, the 10.34% mask accuracy at an IoU of 0.95 indicates the presence of challenging test cases. To further enhance performance, careful removal of problematic training data could be considered, as the dataset was created randomly.

Table 1. mAP comparison between the ground truth and the predicted registered knee image with different IoU thresholds.

| IoU | 0.50 | 0.55 | 0.60 | 0.65 | 0.70 | 0.75 | 0.80 | 0.85 | 0.90 | 0.95 |
|---|---|---|---|---|---|---|---|---|---|---|
| box | 100.0 | 100.0 | 100.0 | 100.0 | 100.0 | 100.0 | 98.96 | 98.87 | 97.59 | 80.33 |
| mask | 100.0 | 100.0 | 100.0 | 99.35 | 99.35 | 98.14 | 98.01 | 96.30 | 86.76 | 10.34 |

When it comes to segmenting speed, the model was able to successfully segment the 500 testing images in 25 seconds. This translates to 20 frames per second, making it potentially suitable for real-time applications such as fluoroscopic videos that typically operate at around 8 frames per second. With its rapid segmenting speed and high accuracy derived from various architectural components and techniques, combined with the utilization of a biomedical dataset, this deep learning model offers a solution for achieving both speed and accuracy in image segmentation tasks.

*3.2 Results of Several Regular Cases*

Figure 4 depicts the original fluoroscopy images along with their corresponding segmented images generated using this deep neural network. The neural network model classifies both PS and CR TKAs (seen in Figure 4a) and predicts the corresponding masks for both femur (red) and tibia (blue), as illustrated in Figure 4b. The masks of both femur and tibia are well- overlaid atop the corresponding silhouettes, as evident from the visualization. The algorithm segmentation outputs are presented as mask images for the femur (Figure 4c) and tibia (Figure 4d), which can be utilized in the subsequent steps for automatically registering 3D CAD models to 2D fluoroscopy images. While there are no quantitative metrics calculated from Figure 4, these qualitative results overall suggest the efficacy of the proposed deep learning approach in accurately segmenting the implants in fluoroscopy images.

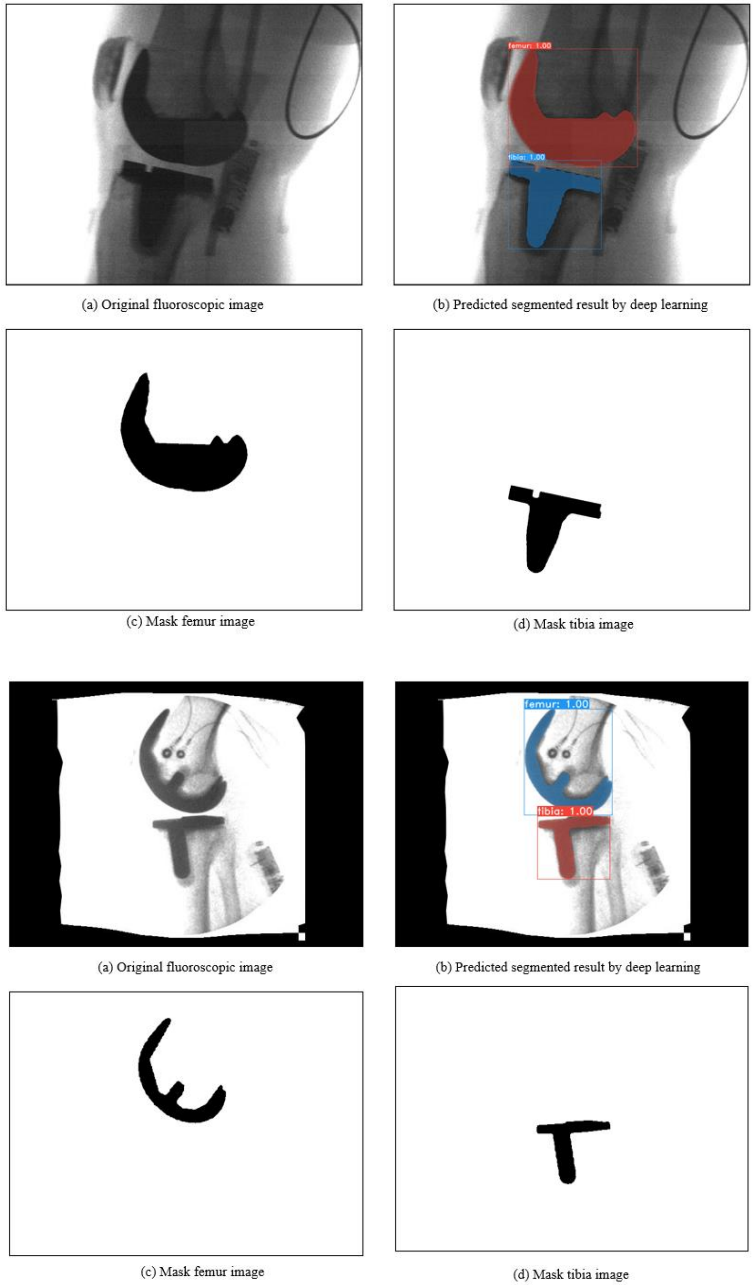

Figure 4. A sample of an original fluoroscopy image of the PS and CR implant patients and the corresponding segmented images.

*3.3 Results of Challenging Cases*

In this section, we present the results focused on challenging cases that can occasionally arise during dynamic data collection, which can pose difficulties for traditional segmentation methods. Figure 5 provides a comprehensive illustration of challenging scenarios in image segmentation, including out-of-captured range cases, overlap cases, patients with both knees implanted, and combination cases involving blurriness and overlap, highlighting the effectiveness of our deep learning model in addressing these complexities. Despite these challenges, the model is able to accurately segment both the femur and tibia, providing evidence of the robustness and effectiveness of our proposed deep learning approach in dealing with challenging images.

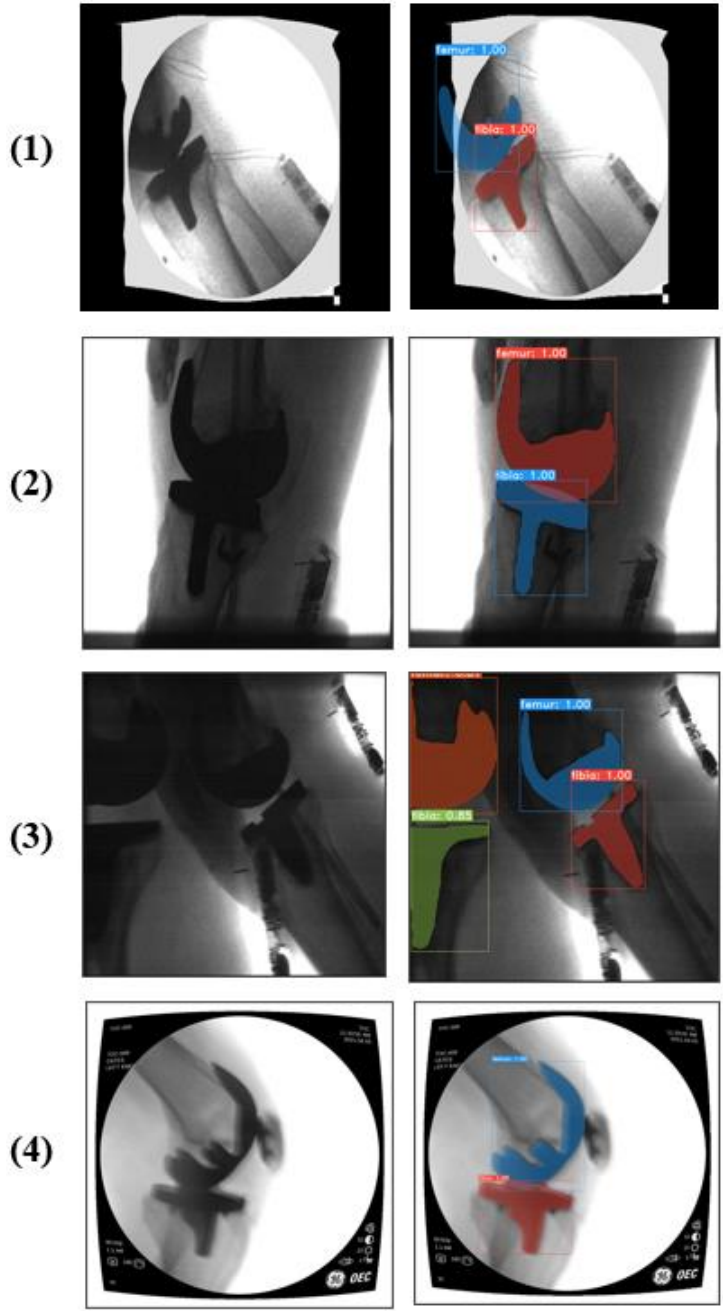

Figure 5. Challenging cases of (1) out-of-captured range, (2) overlap, (3) two implants, (4) combination (blurry and overlap).

*3.4 Comparison with Other Methods*

The qualitative comparisons of the segmented results with manual and thresholding methods are shown in Figure 6. The thresholding method uses the pixel color value to distinguish the darker silhouette of the metallic femur and tibia implants from other components in the fluoroscopic image, resulting in a mask for both implants as shown in the second column of the figure. The manual approach uses Adobe Photoshop to detect the regions of the femur (red) and tibia (blue) separately, which is demonstrated in the third column of the figure. The results obtained through deep learning are presented in the fourth column, where the femur (red) and tibia (blue) are distinguished. In the regular testing cases presented in the figure, the results obtained through deep learning equate to or outperform those achieved using benchmark manual methods.

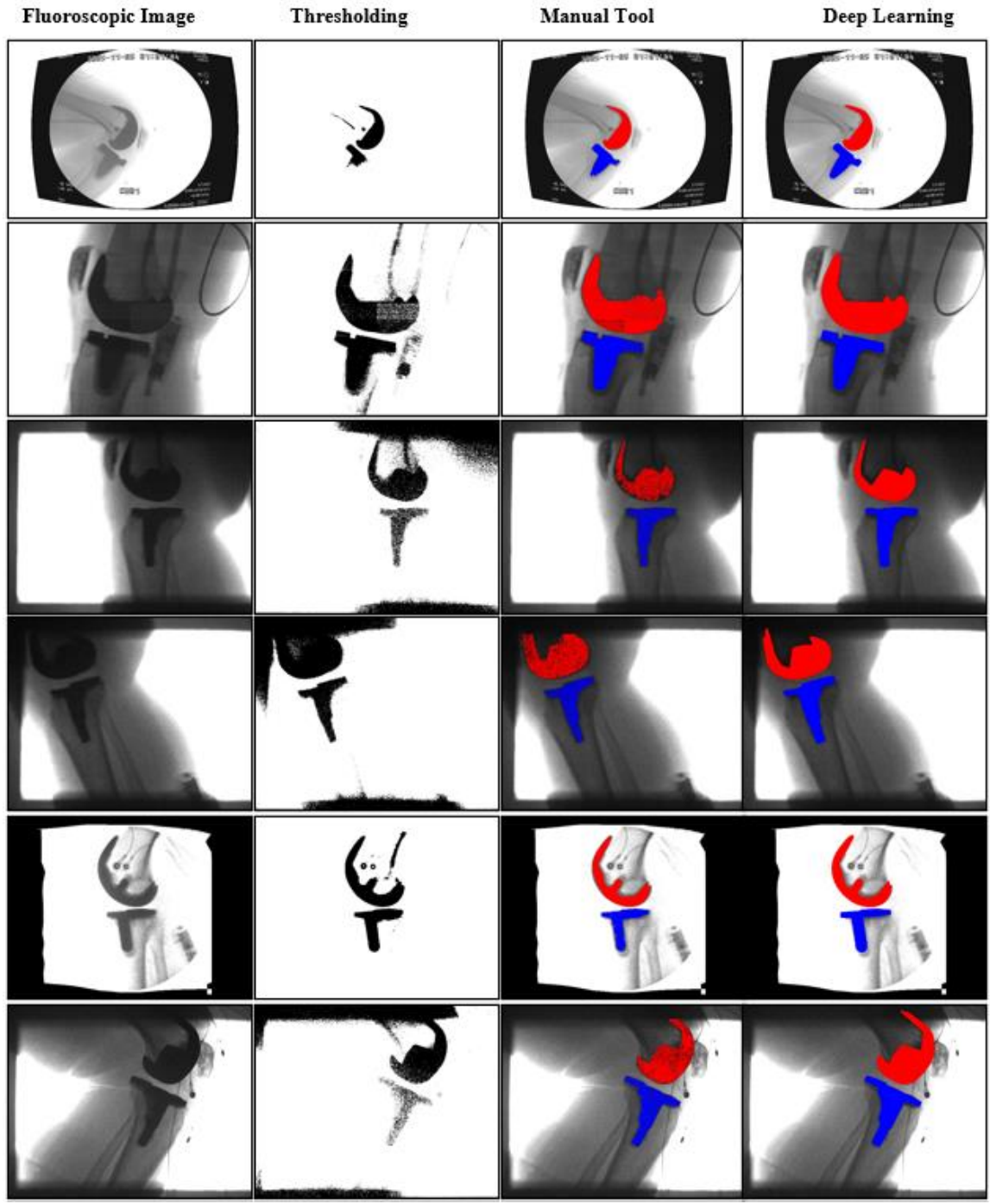

Figure 6. The comparison of the deep learning method with other traditional methods (continued from the previous page).

**4 Discussion**

Automated image segmentation has the potential to revolutionize kinematics analysis by enabling faster and more accurate image analysis. Manual segmentation suffers from drawbacks such as time consumption, human error, and inter-observer variability. In contrast, deep learning-based automated image segmentation offers an objective and reproducible approach for analyzing medical images. This paper utilizes a common orthopaedic dataset and employs deep learning to establish a unique solution for achieving both speed and accuracy in image segmentation tasks within the field of fluoroscopic image analysis.

The various existing methods of automated image segmentation have both strengths and limitations. For example, thresholding [5] is a simple method that is effective for segmenting objects with a well-defined intensity range, but it may not work well for objects with overlapping intensities or inhomogeneous backgrounds. Similarly, edge detection algorithms [6] or statistical shape methods [7] identify boundaries between different structures based on changes in intensity, while clustering algorithms group pixels with similar characteristics into distinct regions [21]. However, those approaches might not work effectively with the low-resolution or overlapping fluoroscopic images. On the other hand, machine learning-based methods can utilize large training data to learn complex image features and can rapidly achieve state-of-the-art performance in various segmentation tasks, as the proposed approach shows a fast and efficient deep neural network model for segmenting implanted TKA fluoroscopy images.

The manual registration approach can also be limited in its ability to register all frames of a fluoroscopy video. In other words, 3D-to-2D fluoroscopic analyses commonly register at intervals of 0, 30, 60, 90, and 120 degrees. While this range can be expanded, it does require more manual work. However, the deep learning model offers a more robust and versatile solution, allowing for segmentation of every frame within a fluoroscopic video. In Figure 7, the machine learning model is demonstrated to be highly robust in segmenting all frames of a fluoroscopy video, meaning this method can capture every single degree of flexion. This robust performance makes the model highly promising for analyzing the kinematics of all frames in a fluoroscopic video and also unlocks the potential for extended real-time applications. Due to the limitation of a visual layout, only 10 degrees increments of flexion are shown in Figure 7.

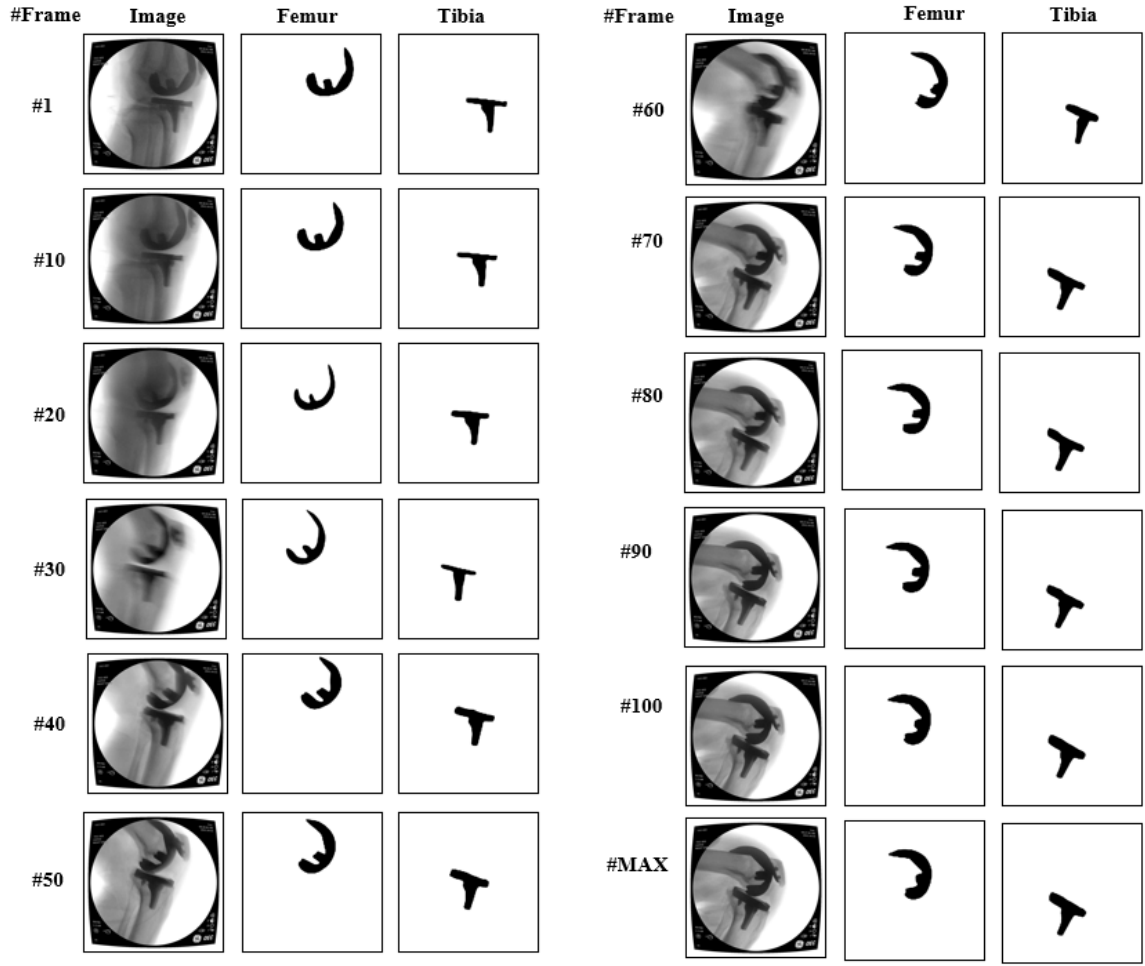

Figure 7. Application of deep learning in segmenting all frames of a fluoroscopic video.

The proposed approach offers significant advantages over other machine learning methods in the field, which utilizes a single deep neural network to generate masks for both femur and tibia and can also successfully segment different implant types such as PS and CR implants. This not only saves time but also reduces the computational power required. The approach is designed to work efficiently on regular computer configurations and average GPUs, making it cost-effective for laboratory and orthopedic room settings.

Additionally, the model exhibits a high processing speed of approximately 20 frames per second, further indicating its potential for real-time use in fluoroscopy. Simulation results demonstrate reliable performance and compare favorably to ground truth and benchmark methods. With its combination of speed and accuracy, this approach holds promise for orthopedic applications.

While deep learning has demonstrated promising results in image segmentation, there are still several limitations to its application. One major challenge is the requirement for large amounts of labeled data to train the model, which can be both time-consuming and expensive and may not be readily available to all groups that are interested. Another limitation is the model's tendency to overfit the data, resulting in poor generalization for new images. Although techniques such as data augmentation can help address this issue, adding more training data has proven to be more effective in improving accuracy. Moreover, deep learning models may struggle with complex structures or variations in lighting, which can lead to segmentation errors. It should also be noted that perfect segmented results cannot always be achieved due to the training dataset containing a variety of implant models from different manufacturers. Despite these limitations, deep learning remains a powerful tool for image segmentation and will continue to be refined and improved through ongoing research and development efforts.

In summary, deep learning-based image segmentation has emerged as a game-changing technique in the field of medical imaging. It can bring significant improvements in accuracy, speed, and robustness compared to traditional methods, with potential applications beyond the knee joint. As medical imaging data continues to grow in quality and quantity, and deep learning algorithms continue to evolve, we can anticipate further breakthroughs and practical solutions for clinical problems. The successful implementation of deep learning-based image segmentation is a critical step in achieving automatic and efficient analysis of 3D implant reconstruction [22], automated 3D-to-2D implant registration, and automated implanted knee kinematics [23]. In the forthcoming work, we will overcome the constraints of the current methodology by pioneering a novel approach aimed at minimizing the expenses associated with utilizing large datasets for training purposes.

# List of References